\documentclass[english,preprint]{revtex4-1}
\usepackage{ae,aecompl}
\usepackage[T1]{fontenc}
\usepackage[latin9]{inputenc}
\usepackage{color}
\usepackage{babel}
\usepackage{mathtools}
\usepackage{bm}
\usepackage{amsmath}
\usepackage{amsthm}
\usepackage{amssymb}
\usepackage{cancel}
\usepackage{mathdots}
\usepackage{stmaryrd}
\usepackage{stackrel}
\PassOptionsToPackage{version=3}{mhchem}
\usepackage{mhchem}
\usepackage{esint}
\usepackage[unicode=true,pdfusetitle,
 bookmarks=true,bookmarksnumbered=false,bookmarksopen=false,
 breaklinks=true,pdfborder={0 0 0},pdfborderstyle={},backref=false,colorlinks=true]
 {hyperref}
\hypersetup{
 urlcolor=red, citecolor=blue}
\usepackage{breakurl}

\makeatletter

\providecommand{\tabularnewline}{\\}

\theoremstyle{remark}
\newtheorem*{rem*}{\protect\remarkname}
\theoremstyle{plain}
\newtheorem{prop}{\protect\propositionname}
\theoremstyle{definition}
 \newtheorem{example}{\protect\examplename}

\usepackage{babel}
\SetSymbolFont{stmry}{bold}{U}{stmry}{b}{n}

\makeatother

\providecommand{\examplename}{Example}
\providecommand{\propositionname}{Proposition}
\providecommand{\remarkname}{Remark}

\begin{document}

\title{Fidelity deviation in quantum teleportation with a two-qubit state }

\author{Arkaprabha Ghosal}
\email{a.ghosal1993@gmail.com}

\affiliation{Centre for Astroparticle Physics and Space Science, Bose Institute,
Block EN, Sector V, Salt Lake, Kolkata 700 091, India}

\author{Debarshi Das}
\email{dasdebarshi90@gmail.com}

\affiliation{Centre for Astroparticle Physics and Space Science, Bose Institute,
Block EN, Sector V, Salt Lake, Kolkata 700 091, India}

\author{Saptarshi Roy}
\email{saptarshiroy@hri.res.in}

\affiliation{Harish-Chandra Research Institute, HBNI, Chhatnag Road, Jhunsi, Allahabad
211 019, India}

\author{Somshubhro Bandyopadhyay}
\email{som@jcbose.ac.in}

\affiliation{Department of Physics and Centre for Astroparticle Physics and Space
Science, Bose Institute, Block EN, Sector V, Salt Lake, Kolkata 700
091, India}
\begin{abstract}
Quantum teleportation with an arbitrary two-qubit state can be appropriately
characterized in terms of maximal fidelity and fidelity deviation.
The former quantifies optimality of the process and is defined as
the maximal average fidelity achievable within the standard protocol
and local unitary strategies, whereas the latter, defined as the standard
deviation of fidelity over all input states, is a measure of fidelity
fluctuations. The maximal fidelity for a two-qubit state is known
and is given by a simple formula that can be exactly computed, but
no such formula is known for the fidelity deviation. In this paper,
we derive an exact computable formula for the fidelity deviation in
optimal quantum teleportation with an arbitrary state of two qubits.
From this formula, we obtain the dispersion-free condition, also known
as the universality condition: the condition that all input states
are teleported equally well and provide a necessary and sufficient
condition for a state to be both useful (maximal fidelity larger than
the classical bound) and universal (zero fidelity deviation). We also
show that for any given maximal fidelity, larger than the classical
bound, there always exist dispersion-free or universal states and
argue that such states are the most desirable ones within the set
of useful states. We illustrate these results with well-known families
of two-qubit states: pure entangled states, Bell-diagonal states,
and subsets of $X$ states.  
\end{abstract}
\maketitle

\section{Introduction}

Quantum teleportation \citep{Teleportation-1993} is a fundamental
protocol to transmit quantum information using shared entanglement
and Local Operations and Classical Communication (LOCC). The novelty
of quantum teleportation lies in the fact that it allows us to transfer
quantum states over quantum channels without physical transmission
of quantum systems. Besides being successfully demonstrated experimentally
\citep{experiment-Boschi,Experiment-Ma,experiment-nolleke}, quantum
teleportation has also been studied in multiparty \citep{Dur-multiparty-teleportation}
and continuous variable systems \citep{CV-Brausnstein}. 

The basic protocol of quantum teleportation is rather simple and can
be understood as an instance of quantum state transfer 
\begin{equation}
\left|\psi\rangle\langle\psi\right|\otimes\rho\overset{{\rm {\rm LOCC}}}{\longrightarrow}\tau\otimes\varsigma,\label{Teleportation-equation}
\end{equation}
where $\rho$ is a two-qubit entangled state initially shared between
the sender (Alice) and the receiver (Bob), $\left|\psi\right\rangle $
is the input state unknown to Alice, $\tau$ is the final state of
Alice's qubits, and $\varsigma$ is the final state of Bob's qubit--the
output state of quantum teleportation. Note that, (\ref{Teleportation-equation})
describes not only the standard protocol \citep{Teleportation-1993}
but also any other protocol within the paradigm of LOCC \citep{MPR-Horodecki-1999,VV-2003}. 

The standard figure of merit for quantum teleportation for a two-qubit
state $\rho$ is given by the average fidelity \citep{RMP-Horodecki-96}
\begin{eqnarray}
\left\langle f_{\rho}\right\rangle  & = & \int f_{\psi,\rho}{\rm d}\psi,\label{teleportationl-fidelity}
\end{eqnarray}
where $f_{\psi,\rho}=\left\langle \psi\left|\varsigma\right|\psi\right\rangle $
is the fidelity between an input-output pair $\left(\left|\psi\rangle\langle\psi\right|,\varsigma\right)$,
and the integral is over a uniform distribution ${\rm d}\psi$ (normalized
Haar measure, $\int{\rm d}\psi=1$) of all input states $\psi=\left|\psi\right\rangle \left\langle \psi\right|$.
The average fidelity, unless mentioned otherwise, is computed with
respect to the standard protocol \citep{Teleportation-1993}. Perfect
teleportation requires $\rho$ to be maximally entangled but this
is possible only when Alice and Bob have access to a noiseless quantum
channel. In practice, however, the available channels are noisy leading
to mixed or noisy entangled states. Consequently, teleportation will
not be perfect and the average fidelity will be less than one. 

Recently \citep{Bang-et-al-2018} it has been pointed out that the
average fidelity alone is not sufficient to fully characterize quantum
teleportation because it does not give us any information on the fluctuations
in fidelity, if any, over the input states, although one expects fluctuations
to be present in general either due to imperfections in experiments,
or even as a property inherent to the resource states. This suggests
that one must also consider, in addition to the average fidelity,
a physically meaningful measure of fluctuations. Fidelity deviation
is one such measure, which is defined as the standard deviation of
fidelity over all input states \citep{Bang-et-al-2012,Bang-et-al-2018}:
\begin{eqnarray}
\delta_{\rho} & = & \sqrt{\left\langle f_{\rho}^{2}\right\rangle -\left\langle f_{\rho}\right\rangle ^{2}},\label{fidelity-standard-deviation}
\end{eqnarray}
where $\left\langle f_{\rho}^{2}\right\rangle =\int f_{\psi,\rho}^{2}{\rm d}\psi$.
Physically, fidelity deviation is a measure of spread of fidelity
values around the average. Note that, 
\begin{eqnarray*}
\delta_{\rho}^{2} & \leq & \left\langle f_{\rho}\right\rangle -\left\langle f_{\rho}\right\rangle ^{2}=\left\langle f_{\rho}\right\rangle \left(1-\left\langle f_{\rho}\right\rangle \right)\leq\frac{1}{4}.
\end{eqnarray*}
Thus $0\leq\delta_{\rho}\leq\frac{1}{2}$, where $\delta_{\rho}=0$
iff $f_{\psi,\rho}=\left\langle f_{\rho}\right\rangle $ for all $\left|\psi\right\rangle $.
Evidently, the ordered pair $\left(\left\langle f_{\rho}\right\rangle ,\delta_{\rho}\right)$
is a more informative performance-measure of quantum teleportation. 

\subsection*{Motivation and Results}

Soon after the original work on quantum teleportation it was realized
that the average fidelity obtained within the standard protocol is
not always optimal \citep{RMP-Horodecki-96}. One can, in fact, maximize
the average fidelity over local unitary (LU) strategies. The maximal
fidelity $F_{\rho}$ is defined as the maximal average fidelity achievable
within the standard protocol and LUs \citep{RMP-Horodecki-96,Badziag-2000},
and any protocol that achieves the maximal value is said to be optimal.
In particular, an optimal protocol is where Alice and Bob first apply
a local unitary operation to transform $\rho$ to the canonical form
$\varrho$ and then use $\varrho$ for quantum teleportation following
the standard protocol \citep{Badziag-2000}. The optimality of the
protocol follows from the equality $F_{\rho}=F_{\varrho}=\left\langle f_{\varrho}\right\rangle $
which shows that the maximal fidelity for a given $\rho$ is the same
as the average fidelity obtained for its canonical representative
$\varrho$ within the standard protocol. The details of this equivalence
and the canonical form will be discussed in the next section. We say
that $\rho$ is useful for quantum teleportation if and only if $F_{\rho}>\frac{2}{3}$
\citep{MPR-Horodecki-1999,Massar_popescu-1995}, where $\frac{2}{3}$
is the maximum average fidelity obtained in classical protocols. 

In this paper, our goal is to obtain the fidelity deviation in optimal
quantum teleportation with an arbitrary two-qubit state $\rho$. Noting
that the maximal fidelity $F_{\rho}$ is realized with the canonical
$\varrho$, we accordingly define the fidelity deviation as 
\begin{eqnarray}
\Delta_{\rho} & = & \delta_{\varrho}.\label{optimal-ordered-pair}
\end{eqnarray}

\begin{rem*}
In general, $\Delta_{\rho}$ is not the same as the minimum of $\delta_{\rho}$,
where the minimum is taken over all LU strategies. This is because
any strategy that maximizes $\left\langle f_{\rho}\right\rangle $
may not minimize $\delta_{\rho}$ and vice versa. But we will see
that there exist states where $\Delta_{\rho}=0$ and for such states
the optimal protocol indeed minimizes the fidelity deviation. Note
that, fidelity deviation is trivially zero for maximally entangled
states as all input states are teleported with fidelity one, but as
we will show, fidelity deviation is nonzero for generic two-qubit
states and states with zero fidelity deviation are in fact special. 
\end{rem*}
Clearly, the ordered pair $\left(F_{\rho},\Delta_{\rho}\right)$ contains
all the necessary information we need to characterize quantum teleportation
with a two-qubit state. But to make use of this performance-measure
we need to know both $F_{\rho}$ and $\Delta_{\rho}$. For an arbitrary
two-qubit state $\rho$, $F_{\rho}$ is given by a computable function
of the eigenvalues of the $T$ matrix, which is a real $3\times3$
matrix with elements $T_{ij}=\text{Tr}\left(\rho\sigma_{i}\otimes\sigma_{j}\right)$,
$i,j=1,2,3$, where $\sigma_{i}$, $i=1,2,3$ are the standard Pauli
matrices \citep{RMP-Horodecki-96,Badziag-2000}. However, no such
formula is known for $\Delta_{\rho}$. In this paper, we derive an
exact computable formula for $\Delta_{\rho}$. This formula is also
function of the eigenvalues of the $T$ matrix. Therefore, knowing
the eigenvalues of $T$ is sufficient to obtain both $F_{\rho}$ and
$\Delta_{\rho}$. 

The main significance of our formula lies in the fact that we can
now adequately characterize quantum teleportation using the ordered
pair $\left(F_{\rho},\Delta_{\rho}\right)$. But there are other useful
implications as well. Because we want fidelity deviation to be as
small as possible, $\Delta_{\rho}$ can serve as a filter to select
the best possible resource states. Here we focus on one such application
-- the universality condition \citep{Bang-et-al-2012,Bang-et-al-2018}
given by $\Delta_{\rho}=0$. So when the universality condition is
satisfied, all input states are teleported equally well. The states
that satisfy the universal condition are said to be universal or dispersion-free
states. From our formula, we obtain the condition under which the
universality condition is satisfied; the analysis reveals that for
a generic $\rho$, $\Delta_{\rho}\neq0$ and that not all useful states
are universal. We then show that a state $\rho$ is both useful and
universal iff for every $i=1,2,3$, $\left|t_{ii}\right|=t>\frac{1}{3}$,
where $t_{ii}$ are the eigenvalues of the $T$ matrix. So if a state
is both useful and universal, then for all input states the fidelity
stays above the classical bound, but the same cannot be said for all
states that are useful but not universal. Hence, the states that are
not only useful but also satisfy the universality condition are the
most desirable ones. 

We now study properties of some of the well-known classes of two-qubit
states. The primary motivation here is to identify which states are
both useful and universal and which are not. The findings are summarized
below: 
\begin{itemize}
\item For a pure entangled state, the fidelity deviation is nonzero unless
the state is maximally entangled. This shows that a nonmaximally entangled
pure state, although useful always \citep{Gisin-1996}, is not universal. 
\item Bell-diagonal states (rank $\geq2$): The motivation to study the
Bell-diagonal states comes from a recent result \citep{Bang-et-al-2018}
which showed that Werner states -- a subset of the Bell-diagonal
states, exhibit zero fidelity deviation. This prompted us to investigate
whether this property extends to all Bell-diagonal states. But we
find that no Bell-diagonal state other than a Werner has zero fidelity
deviation. Thus entangled Werner states are the only states in the
Bell-diagonal family that are both useful and universal. 
\item We present examples of non-Werner states that are shown to be both
useful and universal. These states belong to the family of $X$-states
\citep{X-1,X-2}. 
\end{itemize}
Finally, we would like to remark on a particular property exhibited
by pure entangled states and rank-two entangled Bell-diagonal states.
Both share the common property that fidelity deviation increases with
decrease in the maximal fidelity (equivalently, entanglement for these
states). Now, as entanglement goes to zero we find that the maximal
fidelity approaches the classical bound from above, but the corresponding
fidelity deviation approaches a nonzero constant value from below.
Since the typical region is given by $F_{\rho}\pm\Delta_{\rho}$,
it appears that near the quantum-classical boundary, not all input
states will be teleported with fidelity greater than the classical
bound. In fact, this is one reason why the states with zero fidelity
deviation should be preferred over states with nonzero fidelity deviation,
especially near the quantum-classical boundary. 

The rest of the paper is arranged as follows. Section II reviews the
background material, where we discuss the Hilbert-Schmidt decomposition
and canonical description of two-qubit density matrices. Here we also
summarize the known results on maximal teleportation fidelity. In
Section III, we derive the formula for the fidelity deviation for
a canonical two-qubit state. In Section IV, we discuss the universality
condition and present the relevant case studies. Finally, we conclude
in Section V. 

\section{Preliminaries }

In this section, we review the Hilbert-Schmidt representation and
the canonical form of two-qubit density matrices. We also briefly
discuss maximal teleportation fidelity, present the relevant formulas
and the conditions under which they hold. For more details the readers
are referred to \citep{MPR-Horodecki-1999,RMP-Horodecki-96,RM-Horodecki-1996,Badziag-2000,RP_Horodecki-1996}.

\subsection*{Hilbert-Schmidt representation and the canonical form}

In the Hilbert-Schmidt representation \citep{RMP-Horodecki-96,RP_Horodecki-1996,RM-Horodecki-1996},
a two-qubit density matrix $\rho$ can be written as 

\begin{eqnarray}
\rho & = & \frac{1}{4}\left(I\otimes I+\bm{R}\bm{\cdot\sigma}\otimes I+I\otimes\bm{S}\bm{\cdot\sigma}+\sum_{i,j=1}^{3}T_{ij}\sigma_{i}\otimes\sigma_{j}\right),\label{rho-HSdecomp-1}
\end{eqnarray}
where $\bm{R}$ and $\bm{S}$ are vectors in $\mathbb{R}^{3}$, $\bm{R}\left(\bm{S}\right)\cdot\bm{\sigma}$$=\sum_{k=1}^{3}R_{i}(S_{i})\sigma_{i}$,
and the coefficients $T_{ij}={\rm Tr}\left(\rho\sigma_{i}\otimes\sigma_{j}\right)$,
$i,j=1,2,3$ form a real $3\times3$ matrix $T$ (the correlation
matrix). 

Let $t_{11},t_{22},t_{33}$ be the eigenvalues of the $T$ matrix.
Now one can show that there always exists a product unitary transformation
$U_{1}\otimes U_{2}$ which will transform $\rho\rightarrow\varrho$
\begin{eqnarray}
\varrho & = & \frac{1}{4}\left(I\otimes I+\bm{r\cdot\sigma}\otimes I+I\otimes\bm{s\cdot\sigma}+\sum_{i=1}^{3}\lambda_{i}\left|t_{ii}\right|\sigma_{i}\otimes\sigma_{i}\right),\label{canonical-rho}
\end{eqnarray}
where $\lambda_{i}\in\left\{ -1,+1\right\} $ are determined by the
sign of $\det T$; in particular, (a) if $\det T\leq0$ , then $\lambda_{i}=-1$
for $\left|t_{ii}\right|\neq0$, $i=1,2,3$ ; (b) if $\det T>0$,
then $\lambda_{i},\lambda_{j}=-1$, $\lambda_{k}=+1$ for any choice
of $i\neq j\neq k\in\left\{ 1,2,3\right\} $ satisfying $\left|t_{ii}\right|\geq\left|t_{jj}\right|\geq\left|t_{kk}\right|$.
The transformed state $\varrho$ given by (\ref{canonical-rho}) is
defined as the canonical form of $\rho$ \citep{Badziag-2000}. Note
that this definition, though similar in some ways, differs from that
given in \citep{Badziag-2000}. 

\subsection*{Maximal teleportation fidelity }

For any two-qubit state $\rho$ the maximal fidelity $F_{\rho}$ achievable
within the standard protocol and LUs is given by \citep{MPR-Horodecki-1999}
\begin{eqnarray}
F_{\rho} & = & \frac{2\mathcal{F}_{\rho}+1}{3},\label{max-achievable-fidelity}
\end{eqnarray}
where $\mathcal{F}_{\rho}$ is the fully entangled fraction, defined
as \citep{bennett-1995}
\begin{eqnarray}
\mathcal{F}_{\rho} & = & \max_{\left|\Psi\right\rangle }\left\langle \Psi\left|\rho\right|\Psi\right\rangle .\label{FeF}
\end{eqnarray}
Here, the maximum is taken over all maximally entangled states $\left|\Psi\right\rangle =\left(U\otimes V\right)\left|\Psi_{0}\right\rangle $,
where $\left|\Psi_{0}\right\rangle =\frac{1}{\sqrt{2}}\left(\left|01\right\rangle -\left|10\right\rangle \right)$
is the singlet state and $U$, $V$ are unitary operators. 

From the definition of $\mathcal{F}_{\rho}$ it follows that for any
$\rho^{\prime}=\left(U_{1}\otimes U_{2}\right)\rho\left(U_{1}^{\dagger}\otimes U_{2}^{\dagger}\right)$,
$\mathcal{F}_{\rho}=\mathcal{F}_{\rho^{\prime}}$. Consequently, $F_{\rho}=F_{\rho^{\prime}}$.
Thus to achieve the maximal fidelity, the strategy is to find an appropriate
$\rho^{\prime}$ such that $\mathcal{F}_{\rho^{\prime}}$ is attained
on the singlet state $\left|\Psi_{0}\right\rangle $. In \citep{Badziag-2000}
it was pointed out that a canonical $\varrho$ satisfies this prescription;
that is, $\mathcal{F}_{\varrho}=\left\langle \Psi_{0}\left|\varrho\right|\Psi_{0}\right\rangle $,
which shows that the maximal fidelity $F_{\varrho}$ for a canonical
$\varrho$ is achieved in the standard protocol \citep{Teleportation-1993}.
In other words, the maximal fidelity $F_{\varrho}$ is equal to the
average fidelity $\left\langle f_{\varrho}\right\rangle $. Now using
the fact that $F_{\rho}=F_{\varrho}$ (as $\rho$ and $\varrho$ are
connected by a local unitary transformation), the best strategy for
any given $\rho$ is to first transform $\rho\rightarrow\varrho$
and then use $\varrho$ for quantum teleportation \citep{Badziag-2000}. 

For a canonical $\varrho$ the fully entangled fraction is given by
\citep{Badziag-2000} 
\begin{eqnarray}
\mathcal{F}_{\varrho} & = & \left\{ \begin{array}{ccc}
\frac{1}{4}\left(1+\stackrel[i=1]{3}{\sum}\left|t_{ii}\right|\right) &  & {\rm if}\;\det T\leq0\\
\frac{1}{4}\left[1+\underset{i\neq j\neq k}{\max}\left(\left|t_{ii}\right|+\left|t_{jj}\right|-\left|t_{kk}\right|\right)\right] &  & {\rm if}\;\det T>0
\end{array}\right.,\label{fully-entangled-fraction}
\end{eqnarray}
and using the relation (\ref{max-achievable-fidelity}) one obtains
the maximal fidelity 
\begin{eqnarray}
F_{\varrho} & = & \left\{ \begin{array}{ccc}
\frac{1}{2}\left(1+\frac{1}{3}\stackrel[i=1]{3}{\sum}\left|t_{ii}\right|\right) &  & {\rm if}\;\det T\leq0\\
\frac{1}{2}\left[1+\underset{i\neq j\neq k}{\max}\frac{1}{3}\left(\left|t_{ii}\right|+\left|t_{jj}\right|-\left|t_{kk}\right|\right)\right] &  & {\rm if}\;\det T>0
\end{array}\right..\label{F1andF2}
\end{eqnarray}

\begin{prop}
For a two-qubit state $\rho$, it holds that $F_{\rho}=F_{\varrho}=\left\langle f_{\varrho}\right\rangle $;
$F_{\varrho}$ is given by (\ref{F1andF2}), where $t_{ii}$, $i=1,2,3$
are the eigenvalues of the $T$ matrix associated with $\rho$. 
\end{prop}
The conditions under which a two-qubit state is useful and when it
is not are given by \citep{RMP-Horodecki-96,Badziag-2000}: $F_{\rho}>\frac{2}{3}$
iff $\stackrel[i=1]{3}{\sum}\left|t_{ii}\right|>1$; if $\det T\geq0$,
then $F_{\rho}\leq\frac{2}{3}$. The former implies that all useful
states have the property $\det T<0$; the converse, however, is not
true in general. 
\begin{rem*}
It may so happen that $\rho$ is entangled but $F_{\rho}\leq\frac{2}{3}$.
Such states cannot be directly used for quantum teleportation as they
offer no quantum advantage. However, any such state can be transformed
using appropriate trace-preserving LOCC to another state which is
useful \citep{VV-2003,Badziag-2000,SB-2002}.
\end{rem*}

\section{\label{deviation} Fidelity deviation for a two-qubit state }

In this section, we obtain the fidelity deviation $\Delta_{\rho}$
in optimal quantum teleportation with a two-qubit state $\rho$. Recall
that 
\begin{eqnarray}
\Delta_{\rho} & = & \delta_{\varrho}\nonumber \\
 & = & \sqrt{\left\langle f_{\varrho}^{2}\right\rangle -\left\langle f_{\varrho}\right\rangle ^{2}}.\label{fid-deviation-canonical-formula}
\end{eqnarray}
We now consider the terms appearing in the above formula individually. 

\subsection*{Expression for the average fidelity $\left\langle f_{\varrho}\right\rangle $ }

To obtain an expression for $\left\langle f_{\varrho}\right\rangle $,
we essentially follow the steps given in \citep{RMP-Horodecki-96}
with appropriate modifications and explanations where necessary. 

We assume that Alice and Bob share a canonical two-qubit state $\varrho$.
Alice holds the input qubit prepared in some state $\left|\psi\right\rangle $,
which is unknown to her. First, Alice performs a measurement in the
Bell basis: 
\begin{equation}
\begin{array}{ccc}
\left|\Psi_{0}\right\rangle =\frac{1}{\sqrt{2}}\left(\left|01\right\rangle -\left|10\right\rangle \right) &  & \left|\Psi_{1}\right\rangle =\frac{1}{\sqrt{2}}\left(\left|00\right\rangle -\left|11\right\rangle \right)\\
\left|\Psi_{2}\right\rangle =\frac{1}{\sqrt{2}}\left(\left|00\right\rangle +\left|11\right\rangle \right) &  & \left|\Psi_{3}\right\rangle =\frac{1}{\sqrt{2}}\left(\left|01\right\rangle +\left|10\right\rangle \right)
\end{array}\label{Bell-basis}
\end{equation}
on the two qubits she holds and sends two classical bits to Bob informing
him about the outcome. Subsequently, Bob applies an appropriate Pauli
rotation $\sigma_{k}$, $k\in\left\{ 0,\dots,3\right\} $ to obtain
the output state
\begin{eqnarray}
\varsigma_{k} & = & \frac{1}{p_{k}}{\rm Tr}_{{\rm Alice}}\left[\left(\left|\Psi_{k}\rangle\langle\Psi_{k}\right|\otimes\sigma_{k}\right)\left(\left|\psi\rangle\langle\psi\right|\otimes\varrho\right)\left(\left|\Psi_{k}\rangle\langle\Psi_{k}\right|\otimes\sigma_{k}\right)\right],\label{output-varsigma_k}
\end{eqnarray}
where $p_{k}={\rm Tr}\left[\left(\left|\Psi_{k}\rangle\langle\Psi_{k}\right|\otimes I\right)\left(\left|\psi\rangle\langle\psi\right|\otimes\varrho\right)\right]$
is the probability of the $k^{{\rm th}}$, $k\in\left\{ 0,\dots,3\right\} $
outcome of Alice's measurement and the partial trace is taken over
Alice's qubits. Now using the Bloch sphere representation $\left|\psi\rangle\langle\psi\right|=\frac{1}{2}\left(I+\bm{a\cdot\sigma}\right)$,
where $\bm{a}$ is a unit vector in $\mathbb{R}^{3}$ and the Hilbert-Schmidt
representation of $\varrho$, Eq.$\,$(\ref{output-varsigma_k}) can
be written as 
\begin{eqnarray}
p_{k}\varsigma_{k} & = & \frac{1}{8}\left\{ \left[1+\bm{a}^{T}T_{k}\bm{r}\right]I+O_{k}^{\dagger}\left(\bm{s}+T^{\dagger}T_{k}\bm{a}\right)\bm{\cdot\sigma}\right\} ,\label{p_kvarsigma_k}
\end{eqnarray}
where $T_{k}$s correspond to the projectors $\left|\Psi_{k}\rangle\langle\Psi_{k}\right|$
and are given by: $T_{0}={\rm diag}\left(-1,-1,-1\right)$, $T_{1}={\rm diag}\left(-1,1,1\right)$,
$T_{2}={\rm diag}\left(1,-1,1\right)$, $T_{3}={\rm diag}\left(1,1,-1\right)$;
$\bm{r}$, $\bm{s}$, $T$ correspond to $\varrho$; $O_{k}$s are
rotations in $\mathbb{R}^{3}$ obtained via 
\begin{eqnarray}
\sigma_{k}\left(\bm{n\cdot\sigma}\right)\sigma_{k} & = & \left(O_{k}^{\dagger}\bm{n}\right)\bm{\cdot\sigma}\;\;\;k=0,\dots,3\label{O-defined}
\end{eqnarray}
From (\ref{O-defined}) it then follows that 
\begin{eqnarray}
O_{k}^{\dagger} & = & -T_{k}\;\;\;k=0,\dots,3\label{Okdagger=00003D-Tk}
\end{eqnarray}
Then (\ref{p_kvarsigma_k}) can be written as 
\begin{eqnarray}
p_{k}\varsigma_{k} & = & \frac{1}{8}\left\{ \left[1+\bm{a}^{T}T_{k}\bm{r}\right]I-\left(T_{k}\bm{s}+T^{\dagger}\bm{a}\right)\bm{\cdot\sigma}\right\} ,\label{p_kvarsigma_k-1}
\end{eqnarray}
where we have used the fact that $T_{k}$ commutes with $T^{\dagger}$. 

Now fidelity is given by 
\begin{eqnarray}
f_{\psi,\varrho} & = & \sum_{k=0}^{3}p_{k}\left\langle \psi\left|\varsigma_{k}\right|\psi\right\rangle ={\rm Tr}\left[\left(\sum_{k=0}^{3}p_{k}\varsigma_{k}\right)\left(\left|\psi\rangle\langle\psi\right|\right)\right].\label{fidelity-again}
\end{eqnarray}
Using $\left|\psi\rangle\langle\psi\right|=\frac{1}{2}\left(I+\bm{a\cdot\sigma}\right)$
and (\ref{p_kvarsigma_k-1}), the above equation can be written as
\begin{eqnarray}
f_{\psi,\varrho} & = & \frac{1}{8}\sum_{k=0}^{3}\left[1+\left\{ \bm{a}^{T}T_{k}\left(\bm{r}-\bm{s}\right)-\bm{a}^{T}T\bm{a}\right\} \right].\label{fidelity-O-T-etc}
\end{eqnarray}
For convenience, define the vector $\bm{x}=\bm{r}-\bm{s}$. Then 
\begin{eqnarray}
f_{\psi,\rho} & = & \frac{1}{2}\left(1-\bm{a}^{T}T\bm{a}\right)+\frac{1}{8}\sum_{k=0}^{3}\bm{a}^{T}T_{k}\bm{x}.\label{fidelity-O-T-etc-1}
\end{eqnarray}
The second term in (\ref{fidelity-O-T-etc-1}) vanishes because 

\begin{eqnarray*}
\sum_{k=0}^{3}\bm{a}^{T}T_{k}\bm{x} & = & \bm{a}^{T}\sum_{k=0}^{3}T_{k}\bm{x}\\
 & = & \bm{a}^{T}\left(\sum_{k=0}^{3}T_{k}\right)\bm{x}\\
 & = & 0
\end{eqnarray*}
which follows from the property $\sum_{k=0}^{3}T_{k}=0$. So we have
\begin{eqnarray}
f_{\psi,\varrho} & = & \frac{1}{2}\left(1-\bm{a}^{T}T\bm{a}\right).\label{fidelity-simplified}
\end{eqnarray}
To integrate (\ref{fidelity-simplified}) over all input states we
need Schur's orthogonality lemma on $\mathbb{R}^{d}$ \citep{RMP-Horodecki-96,Bang-et-al-2012}:
For a given group $G$, let $\mathcal{O}_{g}$ be an irreducible orthogonal
representation of any element $g\in G$. Then for every matrix $X$
on $\mathbb{R}^{d}$, 
\begin{eqnarray}
\int_{G}{\rm d}g\mathcal{O}_{g}X\mathcal{O}_{g} & = & \frac{1}{d}{\rm Tr}\left(X\right)I_{d},\label{Schur-1}
\end{eqnarray}
where ${\rm d}g$ is the normalized Haar measure $\left(\int_{G}{\rm d}g=1\right)$
and $I_{d}$ is the identity matrix. In our case, $G$ is the rotation
group $O\left(3\right)$ and the vectors $\bm{a}\in\mathbb{R}^{3}$.
By choosing $\bm{z}$ as the reference unit vector we can write $\bm{a}=R_{a}\bm{z}$,
where $R_{a}$ is the rotation matrix. Then the average over the Bloch
surface is equal to the average over the rotation group, i.e. 
\begin{eqnarray*}
\left\langle f_{\varrho}\right\rangle  & = & \frac{1}{2}\int d\bm{a}\,\left(1-\bm{a}^{T}T\bm{a}\right).
\end{eqnarray*}
Using (\ref{Schur-1}) one obtains $\int d\bm{a}\,\left(\bm{a}^{T}T\bm{a}\right)=\frac{1}{3}{\rm Tr}T$.
Thus 
\begin{eqnarray}
\left\langle f_{\varrho}\right\rangle  & = & \frac{1}{2}\left(1-\frac{1}{3}{\rm Tr}T\right)\label{<frhoc>}
\end{eqnarray}
One can now easily verify that the formulas (\ref{F1andF2}) are indeed
obtained for the conditions $\det T\leq0$ and $\det T>0$. 

\subsection*{Expression for $\left\langle f_{\varrho}^{2}\right\rangle $ }

Our starting point is the expression for the fidelity given by (\ref{fidelity-simplified}).
Squaring both sides 
\begin{eqnarray}
f_{\psi,\varrho}^{2} & = & \frac{1}{4}\left(1-\bm{a}^{T}T\bm{a}\right)^{2}.\label{f^2-1}
\end{eqnarray}
Now the average of $f_{\psi,\varrho}^{2}$ over all possible input
states is the same as the average over all Bloch vectors $\bm{a}$
on the Bloch surface. Therefore, 
\begin{eqnarray}
\left\langle f_{\varrho}^{2}\right\rangle  & = & \frac{1}{4}\int{\rm d}\bm{a}\,\left[\left(1-\bm{a}^{T}T\bm{a}\right)^{2}\right],\label{<f^2>-1}
\end{eqnarray}
where ${\rm d}\bm{a}$ is the normalized Haar measure over the Bloch
surface. Expanding 
\begin{eqnarray}
\left\langle f_{\varrho}^{2}\right\rangle  & = & \frac{1}{4}\int{\rm d}\bm{a}\,\left[1-2\bm{a}^{T}T\bm{a}+\left(\bm{a}^{T}T\bm{a}\right)\left(\bm{a}^{T}T\bm{a}\right)\right].\label{<f^2>-2}
\end{eqnarray}
Now using the identity (see, for e.g \citep{Bang-et-al-2012}) 
\begin{eqnarray*}
\left(\bm{a}^{T}T\bm{a}\right)\left(\bm{a}^{T}T\bm{a}\right) & = & \left(\bm{a}^{T}\otimes\bm{a}^{T}\right)\left(T\otimes T\right)\left(\bm{a}\otimes\bm{a}\right),
\end{eqnarray*}
we can write (\ref{<f^2>-2}) as 
\begin{eqnarray}
\left\langle f_{\varrho}^{2}\right\rangle  & = & \frac{1}{4}\int{\rm d}\bm{a}\,\left[1-2\bm{a}^{T}T\bm{a}+\left(\bm{a}^{T}\otimes\bm{a}^{T}\right)\left(T\otimes T\right)\left(\bm{a}\otimes\bm{a}\right)\right].\label{<f^2>-3}
\end{eqnarray}
To evaluate the above integral we need Schur's lemma on $\mathbb{R}^{d}$
(\ref{Schur-1}) and also the generalization on $\mathbb{R}^{d}\otimes\mathbb{R}^{d}$
\citep{Bang-et-al-2012}: For every matrix $X$ on $\mathbb{R}^{d}\otimes\mathbb{R}^{d}$
\begin{eqnarray}
\int_{G}{\rm d}g\left(\mathcal{O}_{g}\otimes\mathcal{O}_{g}\right)X\left(\mathcal{O}_{g}^{T}\otimes\mathcal{O}_{g}^{T}\right) & = & \mathtt{A}I+\mathtt{B}D+\mathtt{C}P,\label{Schur-2}
\end{eqnarray}
where $P$ is a swap matrix defined by $P\left(\bm{x}_{i}\otimes\bm{x}_{j}\right)=\left(\bm{x}_{j}\otimes\bm{x}_{i}\right)$
and written as $P=\sum_{i,j=0}^{d-1}\left(\bm{x}_{j}\otimes\bm{x}_{i}\right)\left(\bm{x}_{i}\otimes\bm{x}_{j}\right)^{T}$,
$D=\left(\sum_{i=0}^{d-1}\bm{x}_{i}\otimes\bm{x}_{i}\right)\left(\sum_{j=0}^{d-1}\bm{x}_{j}\otimes\bm{x}_{j}\right)$,
where $\left\{ \bm{x}_{i}\right\} $ is an orthonormal basis of $\mathbb{R}^{d}$,
and the coefficients $\mathtt{A},\mathtt{B},\mathtt{C}$ are given
by 
\begin{eqnarray*}
\mathtt{A} & = & \frac{\left(d+1\right){\rm Tr}\left(X\right)-{\rm Tr}\left(XD\right)-{\rm Tr}\left(XP\right)}{d\left(d-1\right)\left(d-2\right)},\\
\mathtt{B} & = & \frac{-{\rm Tr}\left(X\right)+\left(d+1\right){\rm Tr}\left(XD\right)-{\rm Tr}\left(XP\right)}{d\left(d-1\right)\left(d-2\right)},\\
\mathtt{C} & = & \frac{-{\rm Tr}\left(X\right)-{\rm Tr}\left(XD\right)+\left(d+1\right){\rm Tr}\left(XP\right)}{d\left(d-1\right)\left(d-2\right)}.
\end{eqnarray*}
Since in our case $G$ is the rotation group $O\left(3\right)$ and
the vectors $\bm{a}\in\mathbb{R}^{3}$, we can write $\bm{a}=R_{a}\bm{z}$
for some rotation matrix $R_{a}$, where $\bm{z}$ is our reference
unit vector, Then the average over the Bloch surface is equal to the
average over the rotation group. To evaluate (\ref{<f^2>-3}) we consider
the integrals separately. First, 
\begin{eqnarray}
\int{\rm d}\bm{a}\,\left(1-2\bm{a}^{T}T\bm{a}\right) & = & 1-\frac{2}{3}{\rm Tr}T,\label{first-intergal}
\end{eqnarray}
where we have used (\ref{Schur-1}). Next, 
\begin{eqnarray}
\int{\rm d}\bm{a}\,\left(\bm{a}^{T}\otimes\bm{a}^{T}\right)\left(T\otimes T\right)\left(\bm{a}\otimes\bm{a}\right) & = & \frac{1}{15}\left[{\rm Tr}\left(T\otimes T\right)+{\rm Tr}\left(T\otimes TD\right)+{\rm Tr}\left(T\otimes TP\right)\right]\nonumber \\
 & = & \frac{1}{15}\left[\left({\rm Tr}T\right)^{2}+{\rm Tr}\left(TT^{\dagger}\right)+{\rm Tr}T^{2}\right],\label{second integral}
\end{eqnarray}
where we have used (\ref{Schur-2}). Now using (\ref{first-intergal})
and (\ref{second integral}) in (\ref{<f^2>-3}) we arrive at 
\begin{eqnarray}
\left\langle f_{\varrho}^{2}\right\rangle  & = & \frac{1}{4}\left[1-\frac{2}{3}{\rm Tr}T+\frac{1}{15}\left\{ \left({\rm Tr}T\right)^{2}+{\rm Tr}\left(TT^{\dagger}\right)+{\rm Tr}T^{2}\right\} \right].\label{<f^2>as-function-of-O}
\end{eqnarray}

\subsection*{Fidelity deviation $\delta_{\varrho}$ }

From (\ref{<frhoc>}) we get 
\begin{eqnarray}
\left\langle f_{\varrho}\right\rangle ^{2} & = & \frac{1}{4}\left[1-\frac{2}{3}{\rm Tr}T+\frac{1}{9}\left({\rm Tr}T\right)^{2}\right].\label{<f>^2}
\end{eqnarray}
Substituting (\ref{<f^2>as-function-of-O}) and (\ref{<f>^2}) in
the formula (\ref{fid-deviation-canonical-formula}), we finally obtain
\begin{eqnarray}
\delta_{\varrho} & = & \frac{1}{\sqrt{30}}\sqrt{{\rm Tr}T^{2}-\frac{1}{3}\left({\rm Tr}T\right)^{2}},\label{deviation-O}
\end{eqnarray}
which is our desired formula. 
\begin{rem*}
Note that, in the entire derivation the only property of $\varrho$
that we used is that $T$ is diagonal. So the derivation, in fact,
holds for any density matrix with diagonal $T$. 
\end{rem*}
Let us now obtain the explicit expressions for the cases $\det T\leq0$
and $\det T>0$. Recall that for a canonical $\varrho$, $T$ is diagonal
with eigenvalues $\lambda_{i}\left|t_{ii}\right|$, $i=1,2,3$, where
$\lambda_{i}\in\left\{ -1,+1\right\} $. Then Eq.$\,$(\ref{deviation-O})
becomes 
\begin{eqnarray}
\delta_{\varrho} & = & \frac{1}{3\sqrt{10}}\sqrt{2\sum_{i=1}^{3}\left|t_{ii}\right|^{2}-2\sum_{i<j}\lambda_{i}\lambda_{j}\left|t_{ii}\right|\left|t_{jj}\right|}.\label{deviation)-1}
\end{eqnarray}
Now, if $\det T\leq0$, $\lambda_{i}=-1$, $i=1,2,3$, and if $\det T>0$,
$\lambda_{i}=-1,\lambda_{j}=-1,\lambda_{k}=+1$ for any choice of
$i\neq j\neq k\in\left\{ 1,2,3\right\} $ satisfying $\left|t_{ii}\right|\geq\left|t_{jj}\right|\geq\left|t_{kk}\right|$.
Using these properties it is straightforward to arrive at the following
exact expressions: 
\begin{eqnarray}
\delta_{\varrho} & = & \left\{ \begin{array}{ccc}
\frac{1}{3\sqrt{10}}\sqrt{\stackrel[i<j=1]{3}{\sum}\left(\left|t_{ii}\right|-\left|t_{jj}\right|\right)^{2}} &  & {\rm if}\;\det T\leq0\\
\underset{i\neq j\neq k}{\min}\frac{1}{3\sqrt{10}}\sqrt{\left(\left|t_{ii}\right|-\left|t_{jj}\right|\right)^{2}+\left(\left|t_{ii}\right|+\left|t_{kk}\right|\right)^{2}+\left(\left|t_{jj}\right|+\left|t_{kk}\right|\right)^{2}} &  & {\rm if}\;\det T>0
\end{array}\right..\label{delta1and=00005Cdelta2}
\end{eqnarray}
where the minimum taken in the second expression (for $\det T>0$)
is, in fact, obtained for the same set of $i,j,k$, $i\neq j\neq k$
that maximizes the average fidelity given by (\ref{F1andF2}) for
$\det T>0$. This completes the derivation. 

\subsection*{Summary}

The formulas (\ref{delta1and=00005Cdelta2}) and (\ref{F1andF2})
completely characterize quantum teleportation with an arbitrary two-qubit
state $\rho$ within the standard protocol and local unitary operations.
The table below summarizes the expressions for both $F_{\rho}$ and
$\Delta_{\rho}$. 
\begin{center}
{\small{}}%
\begin{tabular}{|c|c|c|}
\hline 
{\small{}$\rho$} & {\small{}$F_{\rho}$} & {\small{}$\Delta_{\rho}$}\tabularnewline
\hline 
\hline 
{\small{}det$T<0$} & {\small{}$\frac{1}{2}\left(1+\frac{1}{3}\stackrel[i=1]{3}{\sum}\left|t_{ii}\right|\right)>\frac{2}{3}\;$iff$\,\stackrel[i=1]{3}{\sum}\left|t_{ii}\right|>1$} & {\small{}$\frac{1}{3\sqrt{10}}\sqrt{\stackrel[i<j=1]{3}{\sum}\left(\left|t_{ii}\right|-\left|t_{jj}\right|\right)^{2}}$}\tabularnewline
\hline 
{\small{}det$T=0$} & {\small{}$\frac{1}{2}\left(1+\frac{1}{3}\stackrel[i=1]{3}{\sum}\left|t_{ii}\right|\right)\leq\frac{2}{3}$} & {\small{}$\frac{1}{3\sqrt{10}}\sqrt{\stackrel[i<j=1]{3}{\sum}\left(\left|t_{ii}\right|-\left|t_{jj}\right|\right)^{2}}$}\tabularnewline
\hline 
{\small{}det$T>0$} & {\small{}$\frac{1}{2}\left[1+\frac{1}{3}\underset{i\neq j\neq k}{\max}\left(\left|t_{ii}\right|+\left|t_{jj}\right|-\left|t_{kk}\right|\right)\right]\leq\frac{2}{3}$} & {\small{}$\underset{i\neq j\neq k}{\min}$$\frac{1}{3\sqrt{10}}\sqrt{\left(\left|t_{ii}\right|-\left|t_{jj}\right|\right)^{2}+\left(\left|t_{ii}\right|+\left|t_{kk}\right|\right)^{2}+\left(\left|t_{jj}\right|+\left|t_{kk}\right|\right)^{2}}$}\tabularnewline
\hline 
\end{tabular}{\scriptsize{} \label{Table 1}}{\scriptsize\par}
\par\end{center}
\begin{rem*}
From the above table we see that for any given $\rho$ we only need
to find the eigenvalues of the $T$ matrix to compute $F_{\rho}$
and $\Delta_{\rho}$. However, as emphasized earlier, the optimal
values are physically realized with the canonical form $\varrho$.
\end{rem*}

\section{The universality condition and Examples}

The universality condition, or equivalently, the condition for zero
fidelity deviation, is given by $\Delta_{\rho}=0$. The states that
satisfy this condition are said to be dispersion-free or universal.
But if a state is not useful for quantum teleportation, then whether
it satisfies the universality condition or not, is irrelevant. So
we focus only on states that are useful for quantum teleportation.
The useful states form a subset of the states with the property $\det T<0$. 

By definition, a state which is useful for quantum teleportation satisfies
\begin{eqnarray}
\sum_{i=1}^{3}\left|t_{ii}\right| & > & 1.\label{useful}
\end{eqnarray}
Now if a useful state is universal, then it also must satisfy
\begin{eqnarray}
\sum_{i<j=1}^{3}\left(\left|t_{ii}\right|-\left|t_{jj}\right|\right)^{2} & = & 0.\label{universal}
\end{eqnarray}
The above condition is fulfilled iff $\left|t_{ii}\right|$, $i=1,2,3$
are all equal. Now we see that the condition $\stackrel[i=1]{3}{\sum}\left|t_{ii}\right|>1$
can hold even if $\left|t_{ii}\right|$ are not all equal. Hence,
a state which is useful for quantum teleportation may not be universal. 
\begin{prop}
Let $\rho$ be a two-qubit state. Then $\rho$ is useful and universal
for quantum teleportation iff for every $i=1,2,3$, $\left|t_{ii}\right|=t>\frac{1}{3}$. 
\end{prop}
The proof follows by noting that $t_{ii}$ are all nonzero (since
$\det T<0$) and that the equations (\ref{useful}) and (\ref{universal})
must be satisfied simultaneously. 
\begin{prop}
Let $S$ be the set of all two-qubit density matrices with maximal
fidelity $F$, where $\frac{2}{3}<F\leq1$. Then there always exist
states in $S$ with the property $\Delta_{\rho}=0$. 
\end{prop}
The states with maximal fidelity $F$ must satisfy the equation 
\begin{eqnarray}
\frac{1}{2}\left(1+\frac{1}{3}\sum_{i=1}^{3}\left|t_{ii}\right|\right) & = & F.\label{prop-1}
\end{eqnarray}
Since $\frac{2}{3}<F\leq1$, the states in $S$ must satisfy $\det T<0$
which implies $t_{ii}$, $i=1,2,3$ must be all nonzero. On the other
hand, $\Delta_{\rho}=0$ implies $\left|t_{ii}\right|$ must all be
equal. Then for a given $F>\frac{2}{3}$ the states with $\Delta_{\rho}=0$
are those with $\left|t_{ii}\right|=2F-1$, $i=1,2,3$. Such states
clearly form a subset of $S$. 

It is rather obvious that for a given fidelity, the most desirable
states are those that are universal and the above proposition shows
such states always exist for any given $F>\frac{2}{3}$. 

We now study the properties of some well-known classes of two-qubit
states. 

\subsection*{Fidelity deviation for pure entangled states}

We know that a two-qubit pure entangled state $\left|\phi\right\rangle $
can be written in the Schmidt form as 
\begin{eqnarray*}
\left|\phi\right\rangle  & = & a\left|0^{\prime}0^{\prime}\right\rangle +b\left|1^{\prime}1^{\prime}\right\rangle ,
\end{eqnarray*}
where $a,b\in\mathbb{R}$, $a\geq b>0$, $a^{2}+b^{2}=1$, and $\left\{ \left|0^{\prime}0^{\prime}\right\rangle ,\left|1^{\prime}1^{\prime}\right\rangle \right\} $
represent the Schmidt basis. The eigenvalues of the $T$ matrix are
given by $t_{11}=2ab$, $t_{22}=-2ab$, $t_{33}=1$. Then the maximal
fidelity is given by 
\begin{eqnarray*}
F_{\phi} & = & \frac{2}{3}\left(1+ab\right)>\frac{2}{3},
\end{eqnarray*}
and the fidelity deviation is given by 
\begin{eqnarray}
\Delta_{\phi} & = & \frac{1}{\sqrt{5}}\left[1-\frac{2}{3}\left(1+ab\right)\right]\label{-}\\
 & = & \frac{1}{\sqrt{5}}\left[1-F_{\phi}\right].\label{Delta_psi-F}
\end{eqnarray}
We see that $\Delta_{\phi}\neq0$ as long as $F_{\phi}\neq1$, i.e.
as long as $\left|\phi\right\rangle $ is not maximally entangled.
So within the class of pure entangled states only maximally entangled
states are both useful and universal for quantum teleportation. 

Now, concurrence \citep{Concurrence-1998} of $\left|\phi\right\rangle $
is given by $C\left(\phi\right)=2ab$. This allows us to write (\ref{-})
as 
\begin{eqnarray}
\Delta_{\phi} & = & \frac{1}{3\sqrt{5}}\left[1-C\left(\phi\right)\right].\label{Delta-psi-C}
\end{eqnarray}
From (\ref{Delta-psi-C}) we see that $\Delta_{\phi}$ decreases with
increase in entanglement. Moreover, from either of the above two expressions
we find that 
\begin{equation}
0\leq\Delta_{\phi}<\frac{1}{3\sqrt{5}}.\label{Delta-psi-bounds}
\end{equation}
The above inequality together with (\ref{-}) show that as $F_{\phi}\rightarrow\frac{2}{3}$
{[}i.e. $C\left(\phi\right)\rightarrow0${]} from above, $\Delta_{\phi}\rightarrow\frac{1}{3\sqrt{5}}$
from below. Therefore, if $F_{\phi}$ is sufficiently close to the
classical bound, then for some input states the fidelity values might
be in the classical domain. 

\subsection*{Fidelity deviation for Bell-diagonal states}

The Bell-basis is defined in (\ref{Bell-basis}). A Bell-diagonal
state $\rho_{{\rm BD}}$ can be written as
\begin{eqnarray}
\rho_{{\rm BD}} & = & \sum_{i=0}^{3}p_{i}\left|\Psi_{i}\right\rangle \left\langle \Psi_{i}\right|,\label{Bell-diagonal}
\end{eqnarray}
where $0\leq p_{3}\leq p_{2}\leq p_{1}<1$ and $\stackrel[i=0]{3}{\sum}p_{i}=1$.
The Hilbert-Schmidt representation of $\rho_{{\rm BD}}$ is given
by 
\begin{eqnarray*}
\rho_{{\rm BD}} & = & \frac{1}{4}\left(I\otimes I+\sum_{i=1}^{3}t_{ii}\sigma_{i}\otimes\sigma_{i}\right),
\end{eqnarray*}
where
\begin{eqnarray}
t_{11} & = & -\left[p_{0}-\left(p_{2}+p_{3}-p_{1}\right)\right]\nonumber \\
t_{22} & = & -\left[p_{0}-\left(p_{1}+p_{3}-p_{2}\right)\right]\label{BD_ts}\\
t_{33} & = & -\left[p_{0}-\left(p_{1}+p_{2}-p_{3}\right)\right]\nonumber 
\end{eqnarray}
Now $\rho_{{\rm BD}}$ is entangled iff $p_{0}>p_{1}+p_{2}+p_{3}$,
or equivalently, $p_{0}>\frac{1}{2}$. Then from (\ref{BD_ts}) it
follows that $t_{11},t_{22},t_{33}<0$. Hence, $\det T<0$. 

Using the formulas for $\Delta_{\rho}$ and $F_{\rho}$ for the states
with $\det T<0$, one obtains 
\begin{eqnarray}
\Delta_{{\rm BD}} & = & \frac{2}{3\sqrt{10}}\sqrt{\left(p_{1}-p_{2}\right)^{2}+\left(p_{2}-p_{3}\right)^{2}+\left(p_{1}-p_{3}\right)^{2}},\label{Delta-BD}
\end{eqnarray}
and 
\begin{eqnarray}
F_{{\rm BD}} & = & \frac{2}{3}\left(\frac{1}{2}+p_{0}\right)>\frac{2}{3}\label{fidelity-BD}
\end{eqnarray}
since $p_{0}>\frac{1}{2}$. 

For rank four states the fidelity deviation is given by (\ref{Delta-BD}).
Thus the states with zero fidelity deviation are those for which $p_{1}=p_{2}=p_{3}=\frac{1-p_{0}}{3}$.
But these states are nothing but Werner states. This is consistent
with the result obtained in \citep{Bang-et-al-2018}.

Now fidelity deviation for rank three states $\left(p_{3}=0\right)$
is given by $\Delta_{{\rm BD}}=\frac{2}{3\sqrt{10}}\sqrt{\left(p_{1}-p_{2}\right)^{2}+p_{2}^{2}+p_{1}^{2}}\neq0$
and that for rank two states $\left(p_{2}=p_{3}=0\right)$ is given
by $\Delta_{{\rm BD}}=\frac{2}{3\sqrt{5}}p_{1}\neq0$. So all rank
two and rank three Bell-diagonal states have nonzero fidelity deviation.
Thus in the Bell-diagonal family only entangled Werner states are
both useful and universal.

Let us now turn our attention to rank two states. For rank two states
we have $\Delta_{{\rm BD}}=\frac{2}{3\sqrt{5}}\left(1-p_{0}\right)$
since $p_{0}+p_{1}=1$, and fidelity is given by (\ref{fidelity-BD}).
Then, as $p_{0}\rightarrow\frac{1}{2}$ from above (i.e. entanglement
goes to zero), $F_{{\rm BD}}\rightarrow\frac{2}{3}$ from above, and
$\Delta_{{\rm BD}}\rightarrow\frac{1}{3\sqrt{5}}$ from below. Thus,
rank-two Bell-diagonal states exhibit the same property exhibited
by pure entangled states near the quantum-classical boundary. 

\subsection*{Non-Werner states with $\Delta_{\rho}=0$ and $F_{\rho}>\frac{2}{3}$ }

In the Hilbert-Schmidt representation, the Bell-diagonal states are
those with local vectors $\bm{R},\bm{S}=0$. So the states that do
not belong to the Bell-diagonal family must have at least one of the
local vectors nonzero. We now give two examples from the class of
$X$-states \citep{X-1,X-2} with the property $\Delta_{\rho}=0$. 
\begin{example}
Consider the family of rank three states: 
\begin{eqnarray*}
\rho & = & p\left|\Psi_{0}\right\rangle \left\langle \Psi_{0}\right|+\frac{1-p}{2}\left(\left|00\right\rangle \left\langle 00\right|+\left|01\right\rangle \left\langle 01\right|\right),\;0<p<1,
\end{eqnarray*}
where $\left|\Psi_{0}\right\rangle $ is the singlet state. One can
easily verify that the states are entangled for all values of $0<p<1$. 

The Hilbert-Schmidt decomposition is given by 
\begin{eqnarray*}
\rho & = & \frac{1}{4}\left[I\otimes I+\left(1-p\right)\sigma_{3}\otimes I-p\sum_{i=1}^{3}\sigma_{i}\otimes\sigma_{i}\right].
\end{eqnarray*}
Note that, $\det T<0$ as all the eigenvalues of the $T$ matrix are
equal to $-p$, where $p>0$. And since the eigenvalues are equal,
$\Delta_{\rho}=0$. The maximal fidelity is given by $F_{\rho}=\frac{1}{2}\left(1+p\right)$.
Thus the states are both useful and universal for quantum teleportation
for all values of $p$, $\frac{1}{3}<p\leq1$ but not useful when
$0<p\leq\frac{1}{3}$, although they are entangled and satisfy the
universality condition. 
\end{example}
The next example has the property that it's both useful and universal
if and only if it's entangled. 
\begin{example}
Consider the family of rank four states: 
\begin{eqnarray*}
\rho & = & p\left|\Psi_{0}\right\rangle \left\langle \Psi_{0}\right|+\frac{1-p}{4}\left(\left|00\right\rangle \left\langle 00\right|+\left|11\right\rangle \left\langle 11\right|\right)+\frac{1-p}{2}\left|01\right\rangle \left\langle 01\right|,\;0<p<1.
\end{eqnarray*}
The Hilbert-Schmidt decomposition is given by 
\begin{eqnarray*}
\rho & = & \frac{1}{4}\left[I\otimes I+\frac{1-p}{2}\left(\sigma_{3}\otimes I-I\otimes\sigma_{3}\right)-p\sum_{i=1}^{3}\sigma_{i}\otimes\sigma_{i}\right].
\end{eqnarray*}
The states are entangled iff $\frac{1}{3}<p\leq1$. Since $t_{11}=t_{22}=t_{33}=-p$,
$\Delta_{\rho}=0$. The maximal fidelity is given by $F_{\rho}=\frac{1}{2}\left(1+p\right)>\frac{2}{3}$
for $p>\frac{1}{3}$. So in this case, the states are both useful
and universal iff they are entangled. 
\end{example}

\section{Discussion and Concluding remarks }

Quantum teleportation with a two-qubit state is generally characterized
by the average fidelity, which tells us how well, on average, we are
able to teleport unknown quantum states. The average fidelity, however,
does not give us any information on the fluctuations in fidelity,
although it is of particular importance to take such fluctuations
into account. Motivated by an earlier work \citep{Bang-et-al-2012},
the authors in \citep{Bang-et-al-2018} considered fidelity deviation,
which is defined as the standard deviation of fidelity over all input
states, as a measure of fidelity-fluctuations. 

For two-qubit states, the maximal fidelity is defined as the maximal
value of the average fidelity achievable within the standard protocol
and local unitary operations. While the maximal fidelity is given
by a simple formula \citep{RMP-Horodecki-96,Badziag-2000}, no such
formula existed for the fidelity deviation. In this paper, we obtained
the formula for the fidelity deviation in optimal quantum teleportation
with an arbitrary two-qubit state. The formula is given by a function
of the eigenvalues of the correlation matrix ($T$ matrix) and can
be effectively computed. Thus for two-qubit states quantum teleportation
can be adequately characterized in terms of maximal fidelity and the
corresponding fidelity deviation. Here we want to emphasize that the
formula for fidelity deviation applies to any two-qubit state, even
if such a state is dynamically generated e.g. through system-bath
interactions (see, for e.g. \citep{open-quantum-1,open-quantum-2,open-quantum-3})
and similar analyses along the lines of our examples are indeed possible. 

Our results showed that fidelity deviation for two-qubit states is
nonzero in general, and the states with zero fidelity deviation are
indeed special. We obtained the condition for zero fidelity deviation
-- the universality condition and provided a necessary and sufficient
condition for a two-qubit state to be both useful and universal. We
also showed that for any given maximal fidelity, which is larger than
the classical bound, there always exist states with zero fidelity
deviation. We also analyzed the properties of some well-known classes
of two-qubit states and found that fidelity deviation for a pure entangled
state is nonzero unless the state is maximally entangled, and in the
Bell-diagonal family, and Werner states are the only states with zero
fidelity deviation. We also provided specific examples of non-Werner
states with zero fidelity deviation -- these states belong to the
family of $X$-states. 

For pure entangled states and rank-two entangled Bell-diagonal states
we found that as entanglement goes to zero, the maximal fidelity approaches
the classical bound from above as we expect, but the corresponding
fidelity deviation approaches a constant value from below. This implies
that in the neighbourhood of the quantum-classical boundary, fidelity
values for some input states may well be in the classical region,
although on average teleportation would still be quantum. Such a situation,
however, doesn't arise for states that are both useful and universal
because all input states are teleported with equal fidelity larger
than the classical bound. This is why universal states form the preferred
subset of all useful states. 

Now one could argue that it should be possible to achieve vanishing
fidelity deviation for any shared two-qubit state by twirling protocols.
For example, before teleportation, Alice applies a random single-qubit
unitary $U$ drawn according to the Haar measure on $SU\left(2\right)$
to the input state, and after teleportation, Bob applies the inverse
$U^{\dagger}$ to the teleported state -- as one effectively averages
over all input states the teleportation fidelity for each state thus
coincides with the average fidelity. Alternatively, they can apply
the random twirl -- a random bilateral unitary operation of the form
$U\otimes U$ (in higher dimensions $U\otimes U^{*}$ \citep{MPR-Horodecki-1999})
-- on the two-qubit resource state (canonical form, or otherwise),
which effectively transforms the two-qubit state to Werner form (note
that they can achieve the same goal by making a random selection,
with uniform probabilities, from a specific set of just twelve unitary
operations $\left\{ U_{i}\right\} $ which involve identical rotations
on each of the two qubits; for details, see, appendices A and B of
\citep{bennett-1995}). Both protocols, however, will lead to a modest
increase in classical communication cost as identity of the unitary
operations must be communicated. Nevertheless, they should be feasible
in practice, although their performance could be limited by experimental
(im)precision of implementation of single qubit unitaries rather than
by the ability to classically transfer specification of the chosen
unitary matrix. 

Let us now discuss in what ways our result could be useful in physical
realizations of quantum teleportation. In experiments, one always
deals with imperfections coming from various sources -- quantum channels,
entangled states, measurements etc., and such imperfections give rise
to fidelity fluctuations. Our formula for the fidelity deviation,
on the other hand, estimates fidelity fluctuations arising from the
given resource state (equivalently, the noisy quantum channel used
to establish such a state). Thus to make use of this formula to analyze
experiments one needs to obtain full information about the state by
tomography before calculating the fidelity deviation as fidelity is
not an observable. Although state tomography with finite samples will
lead to errors, approximate twirling should always be possible with
a finite amount of randomness.

Finally, we expect the results in this paper will help us to understand
the fundamental properties of quantum teleportation and also find
meaningful applications in scenarios where fidelity deviation can
be used as a filter to select optimal resource states. While we have
considered one such application, namely, universality in quantum teleportation,
we think there could be many other possibilities. For example, one
might want to find the optimal states from a given set of states where
the set is well-defined with respect to some physical property. 
\begin{acknowledgments}
DD acknowledges financial support from University Grants Commission
(UGC), Government of India. SB is supported in part by SERB (Science
and Engineering Research Board), Department of Science and Technology,
Government of India through Project No. EMR/2015/002373. 
\end{acknowledgments}


\begin{thebibliography}{99}
\bibitem{Teleportation-1993} C. Bennett, G. Brassard, C. Crepeau,
R. Jozsa, A. Peres, and W. K. Wootters, Teleporting an unknown quantum
state via dual classical and Einstein-Podolsky-Rosen channels, \href{https://journals.aps.org/prl/abstract/10.1103/PhysRevLett.70.1895}{Phys. Rev. Lett. {\bf 70}, 1895 (1993).}

\bibitem{experiment-Boschi} D. Boschi, S. Branca, F. De Martini,
L. Hardy and S. Popescu, Experimental Realization of Teleporting an
Unknown Pure Quantum State via Dual Classical and Einstein-Podolsky-Rosen
Channels, \href{https://journals.aps.org/prl/abstract/10.1103/PhysRevLett.80.1121}{Phys.Rev.Lett. {\bf 80}, 1121 (1998).}

\bibitem{Experiment-Ma} X. Ma, T. Herbst, T. Scheidl, D. Wang, S.
Kropatschek, W. Naylor, B. Wittmann, A. Mech, J. Kofler, E. Anisimova,
V. Makarov, T. Jennewein, R. Ursin and Anton Zeilinger, Quantum teleportation
over 143 kilometres using active feed-forward, \href{https://www.nature.com/articles/nature11472}{Nature {\bf 489}, 269273 (2012).}

\bibitem{experiment-nolleke} C. Nölleke, A. Neuzner, A. Reiserer,
C. Hahn, G. Rempe, and S. Ritter, Efficient Teleportation Between
Remote Single-Atom Quantum Memories, \href{https://journals.aps.org/prl/abstract/10.1103/PhysRevLett.110.140403}{Phys. Rev. Lett. {\bf 110}, 140403 (2013).}

\bibitem{Dur-multiparty-teleportation} W. Dür and J. I. Cirac, Multiparty
teleportation, \href{https://www.tandfonline.com/doi/abs/10.1080/09500340008244039}{Journal of Modern Optics {\bf 47}, 247 (2000).}

\bibitem{CV-Brausnstein} S. L. Braunstein and H. J. Kimble, Teleportation
of Continuous Quantum Variables, \href{https://journals.aps.org/prl/abstract/10.1103/PhysRevLett.80.869}{Phys. Rev. Lett. {\bf 80}, 869 (1998).}

\bibitem{MPR-Horodecki-1999} M. Horodecki, P. Horodecki, and R. Horodecki,
General teleportation channel, singlet fraction, and quasidistillation,
\href{https://journals.aps.org/pra/abstract/10.1103/PhysRevA.60.1888}{Phys. Rev. A {\bf 60}, 1888 (1999).}  

\bibitem{VV-2003} F. Verstraete, and H. Verschelde, Optimal teleportation
with a mixed state of two qubits,  \href{https://journals.aps.org/prl/abstract/10.1103/PhysRevLett.90.097901}{Phys. Rev. Lett. {\bf 90}, 097901 (2003).}

\bibitem{RMP-Horodecki-96} R. Horodecki, M. Horodecki, and P. Horodecki,
Teleportation, Bell's inequalities and inseparability, \href{https://www.sciencedirect.com/science/article/abs/pii/0375960196006391}{Phys. Lett. A {\bf222}, 21 (1996).}

\bibitem{bennett-1995} C. H. Bennett, D. P. DiVincenzo, J. A. Smolin,
and W. K. Wootters, Mixed-state entanglement and quantum error correction,
\href{https://journals.aps.org/pra/abstract/10.1103/PhysRevA.54.3824}{Phys. Rev. A {\bf54}, 3824 (1996).}

\bibitem{Bang-et-al-2012} J. Bang, S.-W. Lee, H. Jeong, and J. Lee,
Procedures for realizing an approximate universal-not gate, \href{https://journals.aps.org/pra/abstract/10.1103/PhysRevA.86.062317}{Phys. Rev. A {\bf 86}, 062317 (2012).}

\bibitem{Bang-et-al-2018} J. Bang, J. Ryu, and D. Kaszlikowski, Fidelity
deviation in quantum teleportation, \href{https://iopscience.iop.org/article/10.1088/1751-8121/aaac35/meta}{J. Phys. A: Math. Theor. {\bf 51}, 135302 (2018).}

\bibitem{RM-Horodecki-1996} R. Horodecki, and M. Horodecki, Information-theoretic
aspects of inseparability of mixed states, \href{https://journals.aps.org/pra/abstract/10.1103/PhysRevA.54.1838}{Phys. Rev. A {\bf 54}, 1838 (1996).}

\bibitem{Badziag-2000} P. Badziag, M. Horodecki, P. Horodecki, and
R. Horodecki, Local environment can enhance fidelity of quantum teleportation,
 \href{https://journals.aps.org/pra/abstract/10.1103/PhysRevA.62.012311}{Phys. Rev. A {\bf 62}, 012311 (2000).}

\bibitem{RP_Horodecki-1996} R. Horodecki, P. Horodecki, Perfect correlations
in the Einstein-Podolsky-Rosen experiment and Bell's inequalities,
\href{https://www.sciencedirect.com/science/article/pii/0375960195009051}{Phys. Lett. A {\bf210}, 227 (1996). }

\bibitem{Gisin-1996} N. Gisin, Nonlocality criteria for quantum teleportation,
\href{https://www.sciencedirect.com/science/article/pii/S0375960196800028}{Phys. Lett. A, {\bf 210}, 157 (1996)}.

\bibitem{Massar_popescu-1995} S. Massar and S. Popescu, Optimal Extraction
of Information from Finite Quantum Ensembles, \href{https://journals.aps.org/prl/abstract/10.1103/PhysRevLett.74.1259}{Phys. Rev. Lett. {\bf 74}, 1259 (1995).}

\bibitem{SB-2002}S. Bandyopadhyay, Origin of noisy states whose teleportation
fidelity can be enhanced through dissipation, \href{https://journals.aps.org/pra/abstract/10.1103/PhysRevA.65.022302}{Phys. Rev. A {\bf 65}, 022302 (2002).}

\bibitem{X-1} T. Yu, J. H. Eberly, Evolution from entanglement to
decoherence of bipartite mixed ``X\textquotedbl{} states, \href{https://dblp.org/db/journals/qic/qic7}{Quantum Inf. Comput. {\bf 7}, 459 (2007).}

\bibitem{X-2} P. E. M. F. Mendonca, M. A. Marchiolli, D. Galetti,
Entanglement universality of two-qubit X-states, \href{https://www.sciencedirect.com/science/article/pii/S000349161400253X}{Ann. Phys. {\bf 351}, 79 (2014).}  

\bibitem{Concurrence-1998}William K. Wootters, Entanglement of Formation
of an Arbitrary State of Two Qubits, \href{https://journals.aps.org/prl/abstract/10.1103/PhysRevLett.80.2245}{Phys. Rev. Lett. {\bf80}, 2245 (1998).}

\bibitem{open-quantum-1} Indranil Chakrabarty, Subhashish Banerjee
and Nana Siddharth, A study of Quantum Correlations in Open Quantum
Systems, \href{https://dl.acm.org/citation.cfm?id=2230916.2230917}{Quantum Information and Computation, \bf{11}, 0541 (2011).}

\bibitem{open-quantum-2} Subhashish Banerjee, V. Ravishankar and
R. Srikanth, Dynamics of entanglement in Two-Qubit Open System Interacting
with a Squeezed Thermal Bath via Dissipative interaction, \href{https://www.sciencedirect.com/science/article/abs/pii/S0003491610000060?via%3Dihub}{Annals of Physics \bf{325} (4), 816--834 (2010).}

\bibitem{open-quantum-3} Subhashish Banerjee, V. Ravishankar and
R. Srikanth, Entanglement Dynamics in Two-Qubit Open System Interacting
with a Squeezed Thermal Bath via Quantum Nondemolition interaction,
\href{https://link.springer.com/article/10.1140%2Fepjd%2Fe2009-00286-2}{ European Physical Journal D \bf{56}, 277 (2010).}
\end{thebibliography}
\end{document}